\documentclass[sigchi-a]{acmart}
\def\BibTeX{{\rm B\kern-.05em{\sc i\kern-.025em b}\kern-.08emT\kern-.1667em\lower.7ex\hbox{E}\kern-.125emX}}
    
\acmDOI{}
\acmConference[The Future of Work @ CHI 2019]{The Future of Work @ CHI 2019}{May 05, 2019}{Glasgow, Scotland, UK}
\acmYear{}
\setcopyright{none}
\acmPrice{}
\acmISBN{}

\begin{document}

%
% The "title" command has an optional parameter, allowing the author to define a "short title" to be used in page headers.
\title{The Future of Skill: What Is It to Be Skilled at Work?}

%
% The "author" command and its associated commands are used to define the authors and their affiliations.
\author{Axel Niklasson}
\affiliation{%
  \institution{Centre for Human-Computer Interaction Design}
  \institution{City, University of London}
}
\email{axel.niklasson@city.ac.uk}

\author{Sean Rintel}
\affiliation{%
  \institution{Microsoft Research}
  \institution{21 Station Road}
  \city{Cambridge}
  \country{UK}}
\email{serintel@microsoft.com}

\author{Stephann Makri}
\affiliation{%
  \institution{Centre for Human-Computer Interaction Design}
  \institution{City, University of London}
}
\email{stephann@city.ac.uk}

\author{Alex Taylor}
\affiliation{%
  \institution{Centre for Human-Computer Interaction Design}
  \institution{City, University of London}
}
\email{alex.taylor@city.ac.uk}
%
% This command processes the author and affiliation and title information and builds
% the first part of the formatted document.
\maketitle

\section{Introduction}
Online collaboration tools claim a position at the centre of modern work.
\begin{quote}
``Slack is the collaboration hub for work, no matter what work you do. It's a place where conversations happen, decisions are made, and information is always at your fingertips.'' \cite{Slack}
\end{quote}
Microsoft Teams asks users to ``Meet the hub for teamwork'', Workplace by Facebook proposes that ``...anything is possible when people work together'' and Cisco WebEx Teams exhort users to ``Make teamwork your best work''. \cite{Microsoft, Facebook, CiscoWebex} Taken as quintessentially modern team working tools, we might imagine these systems to be the future of work.

These tools promise to make collaboration easy by assembling all communication and resources of work in one place. Taking cues from social media, they are targeted at making teamwork more conversational, providing a digital place for those informal “watercooler moments”. To intelligently work without swimming in an ocean of communicative overload, all these services are moving into the heady world of AI, where automation and modelling are set to support decision-making and put the right information at one’s fingertips exactly when needed.

But, what, we should ask, is work and team-work in the ways afforded in tools like Slack and Teams? How exactly is this work modern or indeed the future? What is it to collaborate, whether it be in person or online, in this modern version of work? And how should we understand intelligence here, as something that stands in as a proxy for an intelligent actor or as something that extends and amplifies our own capacities? Needless to say, with a burgeoning market for new collaborative tools and indeed a renewed interest in research on collaboration and the workplace \cite{NationalScienceFoundation}, there are a whole host of questions that invite a much more careful scrutiny of contemporary work practices and the future of work.

In this short paper, we introduce work that is aiming to purposefully venture into this mesh of questions from a different starting point. Interjecting into the conversation, we want to ask: ``what is it to be skilled at work?'' Building on work from scholars like Tim Ingold \cite{Ingold2000}, and strands of longstanding research in workplace studies and CSCW \cite{Goodwin1994, Heath1996}, our interest is in turning the attention to the active work of `being good', or `being skilled', at what we as workers do. As we see it, skill provides a counterpoint to the version of intelligence that appears to be easily blackboxed in systems like Slack, and that ultimately reduces much of what people do to work well together. To put it slightly differently, skill -- as we will argue below -- gives us a way into thinking about work as a much more entangled endeavour, unfolding through multiple and interweaving sets of practices, places, tools and collaborations. In this vein, designing for the future of work seems to be about much more than where work is done or how we might bolt on discrete containers of intelligence. More fruitful would be attending to how we succeed in threading so many entities together to do our jobs well -- in `coming to be skilled'. \footnote{As authors, our interest in skill came about as an intentional counterpoint to the blackboxed ``Skills'' of the Amazon Echo voice assistant and its associated marketing communication, including the mind-bending concept of an ``In-Skill Purchase''.}

These early stages of reflecting on skill are coming to form the basis of the first author's PhD research, that is concerned with the introduction of AI in team collaboration tools and how this affects teams and team members. Having started work in October 2018 he aims to draw on workplace studies and CSCW, and the sociomateriality and practice-oriented theories of Wanda Orlikowski \cite{Orlikowski2000, Orlikowski2007}, to inform an expansive empirical study of collaborative tool use in the workplace.

\section{Skill}
To develop our thinking on skill before venturing into the specifics of collaborative tools, let us for a moment turn our attention to a highly skilled and collaborative practice, surgery. Perceived as inherently skilled, how exactly does a surgeon do his or her job well? To consider this question, we revisit and apply a perspective of skill to some work previously reported by the second author, based on fieldwork in a neurosurgery department at a large hospital in the UK. \cite{Mentis2013}

\begin{marginfigure}
  \includegraphics[width=\marginparwidth]{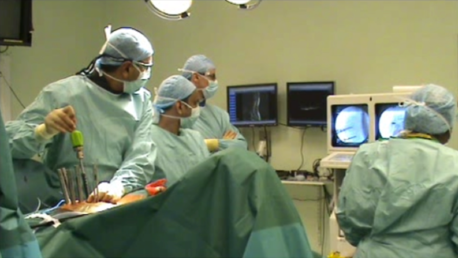}
  \caption{Minimal invasive surgery.}
  \Description{A team of surgeons and medical staff performing minimal invasive surgery in a surgical theatre.}
  \label{fig:surgery}
\end{marginfigure}

One of the vignettes in the report details how something goes amiss in a surgical intervention, during a keyhole procedure (Fig. \ref{fig:surgery}). A wire being threaded into a patient's spine inexplicably fails to grip into the bone. We observe how the surgeon constructs a ``professional vision" \cite{Goodwin1994} of the problem, interweaving his physical manipulation of the wire; the physical response of the patient's body (and spine); iterative viewing of X-ray images and pre-surgery MRI scans; and discussion with his colleagues in the surgical theatre. Altogether and in their unfolding sequence, these make possible what we might refer to as a skilled `vision' for what is going on: the surgeon and his team acknowledge a softness in the bone and agree there are signs of cancer that need to be followed up.

The surgeon (along with his team) is, in the words of Tim Ingold, ``able continually to attune his movements to perturbations in the perceived environment without ever interrupting the flow of action, since that action is itself a process of attention.'' \cite{Ingold2000} This is what sets the skilled practitioner apart from the novice and shows that the surgeon's skill is not an attribute of their person or their job description, the skill is continuously and recursively enacted in the (collaborative) practice.

Taking the surgical theatre as a form of organisational work, skill isn't something easily packaged up. Looking closely at the scenes from the perspective of skill, we find our attention drawn to the following:

\textbf{Skill \textit{distributed} across objects and actors} First, for the team to successfully produce and align their `professional visions' we observed the surgeon's physically manipulation of the wire and the patient's body while iteratively viewing X-ray images requested from a technician. In the indexical requests, the assumed competence of the technician and the repeated enactment of `trial and error' involving the surgeon, the wire, the spine, the technician and the X-ray apparatus we see how skill emerges, distributed across things and actors.

\textbf{Skill made \textit{visible} in the practice} Second, our attention was drawn to the moment the surgeon first perceives a softness of the bone, to the work of aligning what is now seen in the MRI scans with the pre-operation interpretation. In conversation, silent contemplation, in repositioning bodies in the room for perspective and again prodding with the wire we see how the surgeon and his colleagues make skill visible through the practice.

\textbf{Skill \textit{coordinated} across a range of actors} Third, we noted how everyone (and everything) in the room is active in the emerging skill and how the skilled activities are enacted in `full duplex', with the patient's bone responding to the prodding, the technician providing the requested image, a glance returned, a wire held out to grip. In this continuously unfolding sequence of verbal and physical consultation among the surgical team and the theatre we see coordination of skill across a range of actors.

\section{What about Slack?}

\begin{marginfigure}
  \includegraphics[width=\marginparwidth]{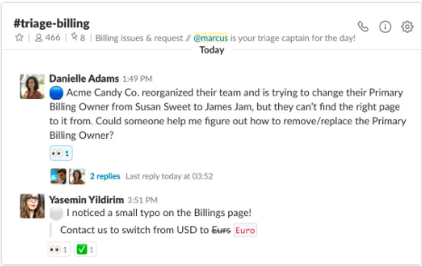}
  \caption{Triage of billing issues in a public Slack channel.}
  \Description{A screenshot of a conversation in a public Slack channel dedicated to triage of billing issues.}
  \label{fig:triage}
\end{marginfigure}

But what of distributed collaborative work? What of the distinctive spatial and temporal discontinuities afforded in the `modern' workplace, and in particular tools like Slack and Microsoft Teams? How might these ideas of skills sit in relation to the corresponding practices and say something about the role of AI? Consider the example of triaging billing issues in a dedicated public Slack channel (Fig. \ref{fig:triage}), a ``success strategy'' suggested by the developers of Slack \cite{Slacka}. Although, admittedly a slight leap from our surgery example, what we want to show is that skill becomes a helpful ``tool for thinking''. \cite{Stengers2005}

First, take the notion of skill distributed across objects and actors. What we begin to see is that the structure of communication in the tool and what it enables and authorises is an integral part of being skilled as a team. The sequentiality of answers to questions, their timeliness, their indexical reference to prior turns and other resources, and so on, are made possible through the tool and become ways in which skill is conducted. If we are to imagine the impact of AI on the skilled work of teams, on how things get done, what might then a modern and future version of work and team-work be, and what are the material consequences?

Second, with the notion of skill made visible in the practice we start to discover how skilled activity flows through, along and around the collaborative tool, how workers are invited to participate in and get notified of activity, how they participate in and choose to withdraw from activity and how documents and activity are enabled to coexist on- and offline. In the conception of AI contributing to the perception of skill-in-practice, what is it to collaborate, whether it be in person or online, and what new forms and notions of skill might then emerge?

\begin{marginfigure}
  \includegraphics[width=\marginparwidth]{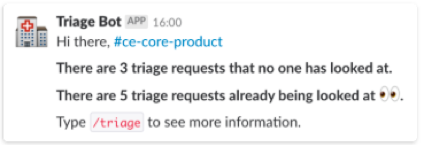}
  \caption{A bot automatically keeping track of triage requests.}
  \Description{A screenshot of communication from a Slack chatbot automatically keeping track of triage requests.}
  \label{fig:bot}
\end{marginfigure}

Third, the notion of skill coordinated across a range of actors opens up for an examination of skilled activity not only across human but also non-human actors, external dependencies and internal analysis of interaction (Fig. \ref{fig:bot}). In an attempt to envisage AI's role as either a proxy for an intelligent actor or as something that extends and amplifies our own capacities, how are our conventional considerations of materiality reconfigured to understand the shifting sociomaterial assemblages involved in modern and future organising?

\section{Conclusions}

In recognition of work such as Orlikowski's \cite{Orlikowski2000} and her theorising of the ``practice lens'', our intention here is not to formalise a general `model' of skill to be applied across different practices. We do however hope to have shown skill as a useful starting point for the study of collaboration at work and how we could think differently about the role for AI. In the workshop, our hope would be to, i., set out this thinking with skill as a resource for further discussion, and, ii., put skill in conversation with some of the broader questions of global labour, \cite{Irani2013a, Lindtner2014} and how it is we come to judge what counts as modern (team) work and how its differentiated and uneven prescriptions might matter for setting an agenda for the future of work.

\section{Author Biographies}

\textbf{Axel Niklasson} is a PhD student at the Centre for Human-Computer Interaction Design (HCID), City, University of London. His current research is centred on the introduction of AI in team collaboration. Previous to his academic work Axel enjoyed a 20-year international career in telecom, media and software engineering. He earned his MSc from Lund University, Sweden. \\ \\
\textbf{Sean Rintel} is a Researcher at the Human Experience \& Design (HXD) group at Microsoft Research Cambridge (UK). His research explores communication technologies for the future of work. He is currently interested in inclusive design for hybrid meetings and video-mediated collaboration, AI in enterprise communication and the gig economy, and AI for mobile functional images. He has been a past CHI AC, past Senior Editor for Communication Technology of the Oxford Research Encyclopedia of Communication, and past Chair of Electronic Frontiers Australia. \\ \\
\textbf{Stephann Makri} is a Senior Lecturer in Human-Computer Interaction (HCI), at the intersection of HCI and Information Science. He specialises in Human Information Interaction (how people interact holistically with digital information environments - e.g. websites, search engines, digital libraries, social media). His work involves understanding peoples' experiences of actively seeking, passively encountering, interpreting and using digital information and using this understanding to design new and improve existing digital information environments. \\ \\
\textbf{Alex Taylor} is a sociologist at the Centre for HCID. With a fascination for the entanglements between social life and machines, his research ranges from empirical studies of technology in everyday life to speculative design interventions. He draws on feminist technoscience to ask questions about the roles human-machine composites play in forms of knowing and being, and how they open up possibilities for fundamental transformations in society.

%
% The acknowledgments section is defined using the "acks" environment (and NOT an unnumbered section). This ensures
% the proper identification of the section in the article metadata, and the consistent spelling of the heading.
\begin{acks}
Many thanks to Simone Stumpf and Adrian Bussone for invaluable feedback. This work was supported by Microsoft Research through its PhD Scholarship Programme.
\end{acks}

%
% The next two lines define the bibliography style to be used, and the bibliography file.
\bibliographystyle{ACM-Reference-Format}
\bibliography{futureofskill}


\begin{thebibliography}{15}

%%% the Web address.

\ifx \showCODEN    \undefined \def \showCODEN     #1{\unskip}     \fi
\ifx \showDOI      \undefined \def \showDOI       #1{#1}\fi
\ifx \showISBNx    \undefined \def \showISBNx     #1{\unskip}     \fi
\ifx \showISBNxiii \undefined \def \showISBNxiii  #1{\unskip}     \fi
\ifx \showISSN     \undefined \def \showISSN      #1{\unskip}     \fi
\ifx \showLCCN     \undefined \def \showLCCN      #1{\unskip}     \fi
\ifx \shownote     \undefined \def \shownote      #1{#1}          \fi
\ifx \showarticletitle \undefined \def \showarticletitle #1{#1}   \fi
\ifx \showURL      \undefined \def \showURL       {\relax}        \fi
% invisible to TeX
\providecommand\bibfield[2]{#2}
\providecommand\bibinfo[2]{#2}
\providecommand\natexlab[1]{#1}
\providecommand\showeprint[2][]{arXiv:#2}

\bibitem[\protect\citeauthoryear{{Cisco Webex}}{{Cisco Webex}}{2019}]%
        {CiscoWebex}
\bibfield{author}{\bibinfo{person}{{Cisco Webex}}.} \bibinfo{year}{2019}\natexlab{}.
\newblock \bibinfo{title}{{Team Collaboration App, File Sharing, Messaging}}.
\newblock
\newblock
\urldef\tempurl%
\url{https://www.webex.com/team-collaboration.html}
\showURL{%
\tempurl}


\bibitem[\protect\citeauthoryear{Facebook}{Facebook}{2019}]%
        {Facebook}
\bibfield{author}{\bibinfo{person}{Facebook}.} \bibinfo{year}{2019}\natexlab{}.
\newblock \bibinfo{title}{{Workplace by Facebook: A work collaboration tool}}.
\newblock
\newblock
\urldef\tempurl%
\url{https://www.facebook.com/workplace}
\showURL{%
\tempurl}


\bibitem[\protect\citeauthoryear{Goodwin}{Goodwin}{1994}]%
        {Goodwin1994}
\bibfield{author}{\bibinfo{person}{Charles Goodwin}.} \bibinfo{year}{1994}\natexlab{}.
\newblock \showarticletitle{{Professional Vision}}.
\newblock \bibinfo{journal}{\emph{American Anthropologist}} \bibinfo{volume}{96}, \bibinfo{number}{3} (\bibinfo{year}{1994}), \bibinfo{pages}{606--633}.
\newblock
\urldef\tempurl%
\url{https://www.jstor.org/stable/682303}
\showURL{%
\tempurl}


\bibitem[\protect\citeauthoryear{Heath and Luff}{Heath and Luff}{1996}]%
        {Heath1996}
\bibfield{author}{\bibinfo{person}{Christian Heath} {and} \bibinfo{person}{Paul Luff}.} \bibinfo{year}{1996}\natexlab{}.
\newblock \showarticletitle{{Documents and professional practice: ‘bad' organisational reasons for ‘good' clinical records}}.
\newblock \bibinfo{journal}{\emph{Computer Supported Cooperative Work (CSCW)}} (\bibinfo{year}{1996}), \bibinfo{pages}{354--363}.
\newblock
\showISBNx{0897917650}
\urldef\tempurl%
\url{https://doi.org/10.1145/240080.240342}
\showDOI{\tempurl}


\bibitem[\protect\citeauthoryear{Ingold}{Ingold}{2000}]%
        {Ingold2000}
\bibfield{author}{\bibinfo{person}{Tim Ingold}.} \bibinfo{year}{2000}\natexlab{}.
\newblock \bibinfo{booktitle}{\emph{{The Perception of the Environment: Essays in Livelihood, Dwelling, and Skill}}}.
\newblock 465 pages.
\newblock
\showISBNx{0203466020}
\showISSN{1097-3729}
\urldef\tempurl%
\url{https://doi.org/10.1353/tech.2002.0079}
\showDOI{\tempurl}
\showeprint[arxiv]{arXiv:1011.1669v3}


\bibitem[\protect\citeauthoryear{Irani and Silberman}{Irani and Silberman}{2013}]%
        {Irani2013a}
\bibfield{author}{\bibinfo{person}{Lilly C~Lc Irani} {and} \bibinfo{person}{M.~Six Silberman}.} \bibinfo{year}{2013}\natexlab{}.
\newblock \showarticletitle{{Turkopticon: Interrupting Worker Invisibility in Amazon Mechanical Turk}}. In \bibinfo{booktitle}{\emph{Proceedings of the SIGCHI Conference on Human Factors in Computing Systems}}. \bibinfo{pages}{611--620}.
\newblock
\showISBNx{9781450318990}
\showISSN{2415-0762}
\urldef\tempurl%
\url{https://doi.org/10.1145/2470654.2470742}
\showDOI{\tempurl}


\bibitem[\protect\citeauthoryear{Lindtner, Hertz, and Dourish}{Lindtner et~al\mbox{.}}{2014}]%
        {Lindtner2014}
\bibfield{author}{\bibinfo{person}{Silvia Lindtner}, \bibinfo{person}{Garnet Hertz}, {and} \bibinfo{person}{Paul Dourish}.} \bibinfo{year}{2014}\natexlab{}.
\newblock \showarticletitle{{Emerging Sites of HCI Innovation: Hackerspaces, Hardware Startups {\&} Incubators}}. In \bibinfo{booktitle}{\emph{Proceedings of the SIGCHI Conference on Human Factors in Computing Systems}}. \bibinfo{pages}{439--448}.
\newblock
\showISBNx{9781450324731}
\urldef\tempurl%
\url{http://dx.doi.org/10.1145.2556288.2557132}
\showURL{%
\tempurl}


\bibitem[\protect\citeauthoryear{Mentis and Taylor}{Mentis and Taylor}{2013}]%
        {Mentis2013}
\bibfield{author}{\bibinfo{person}{Helena~M. Mentis} {and} \bibinfo{person}{Alex~S. Taylor}.} \bibinfo{year}{2013}\natexlab{}.
\newblock \showarticletitle{{Imaging the body: embodied vision in minimally invasive surgery}}. In \bibinfo{booktitle}{\emph{Proceedings of the SIGCHI Conference on Human Factors in Computing Systems}}. \bibinfo{pages}{1479--1488}.
\newblock
\showISBNx{978-1-4503-1899-0}
\urldef\tempurl%
\url{https://doi.org/10.1145/2466110.2466197}
\showDOI{\tempurl}


\bibitem[\protect\citeauthoryear{Microsoft}{Microsoft}{2019}]%
        {Microsoft}
\bibfield{author}{\bibinfo{person}{Microsoft}.} \bibinfo{year}{2019}\natexlab{}.
\newblock \bibinfo{title}{{Microsoft Teams – Group Chat software}}.
\newblock
\newblock
\urldef\tempurl%
\url{https://products.office.com/en-US/microsoft-teams/group-chat-software}
\showURL{%
\tempurl}


\bibitem[\protect\citeauthoryear{{National Science Foundation}}{{National Science Foundation}}{2019}]%
        {NationalScienceFoundation}
\bibfield{author}{\bibinfo{person}{{National Science Foundation}}.} \bibinfo{year}{2019}\natexlab{}.
\newblock \bibinfo{title}{{Future of Work at the Human-Technology Frontier: Core Research}}.
\newblock
\newblock
\urldef\tempurl%
\url{https://www.nsf.gov/pubs/2019/nsf19541/nsf19541.htm}
\showURL{%
\tempurl}


\bibitem[\protect\citeauthoryear{Orlikowski}{Orlikowski}{2000}]%
        {Orlikowski2000}
\bibfield{author}{\bibinfo{person}{Wanda~J. Orlikowski}.} \bibinfo{year}{2000}\natexlab{}.
\newblock \showarticletitle{{Using Technology and Constituting Structures: A Practice Lens for Studying Technology in Organizations}}.
\newblock \bibinfo{journal}{\emph{Organization Science}} \bibinfo{volume}{11}, \bibinfo{number}{4} (\bibinfo{year}{2000}), \bibinfo{pages}{404--428}.
\newblock
\showISBNx{10477039}
\showISSN{1047-7039}
\urldef\tempurl%
\url{https://doi.org/10.1287/orsc.11.4.404.14600}
\showDOI{\tempurl}


\bibitem[\protect\citeauthoryear{Orlikowski}{Orlikowski}{2007}]%
        {Orlikowski2007}
\bibfield{author}{\bibinfo{person}{Wanda~J. Orlikowski}.} \bibinfo{year}{2007}\natexlab{}.
\newblock \showarticletitle{{Sociomaterial practices: Exploring technology at work}}.
\newblock \bibinfo{journal}{\emph{Organization Studies}} \bibinfo{volume}{28}, \bibinfo{number}{9} (\bibinfo{year}{2007}), \bibinfo{pages}{1435--1448}.
\newblock
\showISBNx{0170-8406}
\showISSN{01708406}
\urldef\tempurl%
\url{https://doi.org/10.1177/0170840607081138}
\showDOI{\tempurl}


\bibitem[\protect\citeauthoryear{Slack}{Slack}{2019a}]%
        {Slacka}
\bibfield{author}{\bibinfo{person}{Slack}.} \bibinfo{year}{2019}\natexlab{a}.
\newblock \bibinfo{title}{{Prioritize tasks quickly with triage channels – Slack Help Center}}.
\newblock
\newblock
\urldef\tempurl%
\url{https://get.slack.help/hc/en-us/articles/360000384726-Prioritize-}
\showURL{%
\tempurl}


\bibitem[\protect\citeauthoryear{Slack}{Slack}{2019b}]%
        {Slack}
\bibfield{author}{\bibinfo{person}{Slack}.} \bibinfo{year}{2019}\natexlab{b}.
\newblock \bibinfo{title}{{Where work happens | Slack}}.
\newblock
\newblock
\urldef\tempurl%
\url{https://slack.com/}
\showURL{%
\tempurl}


\bibitem[\protect\citeauthoryear{Stengers}{Stengers}{2005}]%
        {Stengers2005}
\bibfield{author}{\bibinfo{person}{Isabelle Stengers}.} \bibinfo{year}{2005}\natexlab{}.
\newblock \showarticletitle{{Introductory Notes on an Ecology of Practices}}.
\newblock \bibinfo{journal}{\emph{Cultural Studies Review}} \bibinfo{volume}{11}, \bibinfo{number}{1} (\bibinfo{year}{2005}), \bibinfo{pages}{183--196}.
\newblock
\showISBNx{1111111111}
\showISSN{1837-8692}
\urldef\tempurl%
\url{https://doi.org/10.4324/9781315078397}
\showDOI{\tempurl}


\end{thebibliography}

\end{document}